# Revival of the *Silk Road* using the applications of AR/VR and its role on cultural tourism

**Sahar Zandi[1]**

[1] Department of Architecture, Texas A&M University, College Station, TX 77845, USA

**Abstract:**
This research project seeks to investigate the incorporation of augmented reality (AR) and virtual reality (VR) technology with human-computer interaction (HCI) in order to revitalize the Silk Road - specifically in Kermanshah, Iran - and its effect on cultural tourism. Kermanshah has underexplored the rich historical significance of the Silk Road, despite the presence of 24 UNESCO World Heritage sites. From the 2nd century BCE to the 18th century CE, the Silk Road was a vital trade route connecting the West and the East and had enormous cultural, economic, religious, and political effects. The purpose of this study is to examine the application of AR/VR technologies in HCI for the preservation, interpretation, and promotion of the Silk Road's tangible and intangible cultural heritage in Kermanshah, as well as their impact on cultural tourism development. The study also investigates how these innovative technologies can enhance visitors' experiences through immersive and interactive approaches, promote sustainable tourism practices, and contribute to the region's broader socioeconomic benefits. The research will analyze the challenges and opportunities of implementing AR/VR technology in HCI within the context of cultural heritage and tourism in Kermanshah and the Silk Road region more broadly. By combining HCI, AR/VR, and cultural tourism, this research seeks to provide valuable insights into the development of user-centered, immersive experiences that promote a deeper understanding and appreciation of the Silk Road's distinctive cultural heritage.

# 1 INTRODUCTION

## 1.1 Background of the Study

The history of Iran dates back to the Mesopotamian civilization when the Jesus had not been born. With regard to this issue as well as other historical eras like the Islamic Empire and the Persian Empire, Iran accounts for 1 Natural Heritage site and 24 UNESCO Cultural Heritage sites, whereas its expansive national lands and climate involve other tourist sites and tourism resources like geo parks, ski resorts, and deserts. Accordingly, Persian heritage has been researched extensively, but despite the fact that heritage tourism is an attractive topic for academic research, there are research gaps that deserve further attention. In particular, Kermanshah has an important and deep connection to the Silk Road, and the legacy of the historical Silk Road still exists there. However, there have been very few studies exploring Silk Road heritage in Kermanshah city. Kermanshah was therefore selected as the location for the current study of heritage tourism in relation to the Silk Road. It should be mentioned that the Silk Road has been and is one of the networks of trade routes that connects the West and East, and has been very crucial to the cultural, economic, religious as well as political interactions between the mentioned regions from the second century BCE to the 18$^{th}$ century, Fig. 1. In fact, the Silk Road includes the land but also sea routes that link Southeast Asia and East Asia with Persia, South Asia, East Africa, Southern Europe, and the Arabian Peninsula (Elisseeff, 2000).

In this way, experts in the field exchanged techniques and commodities/goods along the Silk Road and thus there was a stream of monks and priests that moved into the East bringing along Zoroastrianism, Manichaeism, Nestorian Christianity, Buddhism, and consequently Islam (Toniolo et al., 2012) .Moreover, International trading has been essential for Sassanid Persians so that the important routes throughout Iran developed fundamentally with the beginning of the first century AD. In addition, the "imperial road" branch, which started in Herat (that is now in Afghanistan) led northward to Merv and then to Samarkand wherein this road probably merged with the Silk Road from China across the oases of Eastern Turkestan. Furthermore, Syria and the area of Asia Minor was linked with the Silk Road through an overland road leading along the Euphrates to the Persian Gulf's harbors, or through an ancient caravan route from Syria in Iran. The commonest things were Chinese raw silk, luxury goods, as well as Indian commodities like aromas, jewels, spices and opium, which were delivered to Iran basically by land. Sassanid Persians periodically carried on wars with Byzantium in their struggles to dominate the busiest sections of the Silk Road. For example, a significant part of the Silk Road, from the borderlines of China to Iran and the trade routes, which led to Western and Southern Siberia, were controlled by Turkic Khaganate in the 5th century. Due to the conflicts for the amount of silk trade, Turkic people came into conflict with Iran and attempted to construct a novel trade route to Byzantium, which bypassed the steppes of the Black Sea coast and Iran across the Volga area (web, 1).

Even though the Silk Road contributed importantly to the trade, culture, as well as religious exchange between different regions, "Silk Road" did not appear in historical research until the late 19th century. Attention paid to the Silk Road in relation to tourism development was initiated later in the 1990s when the United Nations World Tourism Organization (UNWTO) called for its revitalization through cultural conservation and tourism (Whitfield, 2007).

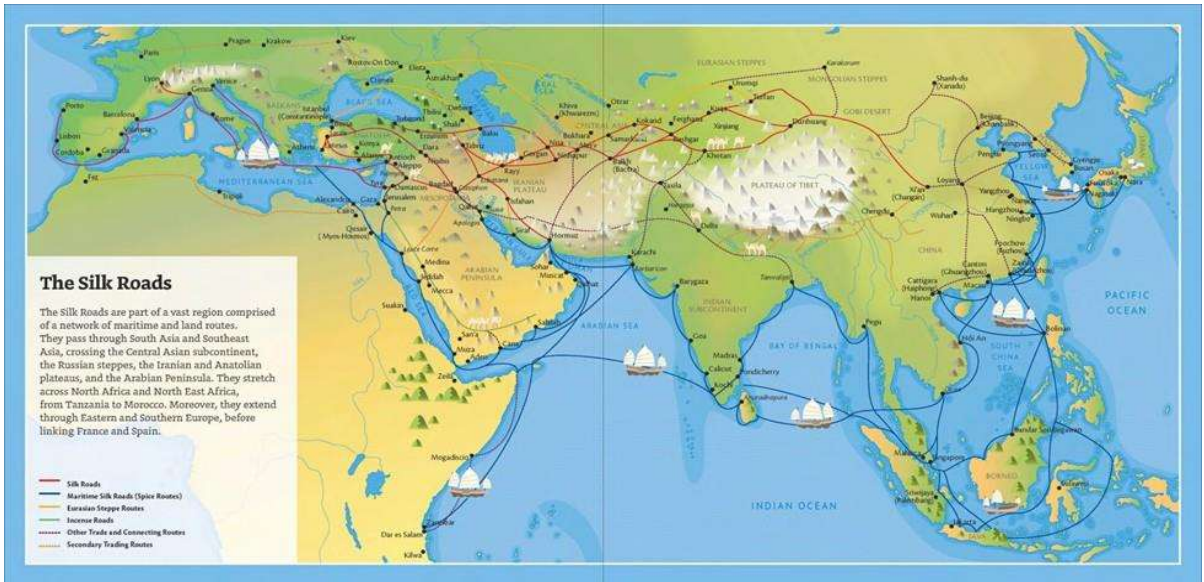

Figure 1: Map of the Silk Road (web, 2).

## 1.2  Research Objective and Questions

This study aims to understand the relationship between the Silk Road and heritage tourism in Iran and preserve the historic places along this ancient Road. The research aims at answering these questions:

1) What is the current status of the Silk Road heritage in Iran?
2) Can the Silk R be useful as a tourist attraction in Iran?
3) What can be done to conserve historic places along this ancient Road that are difficult to reconstruct? (my main question in research)

## 1.3  Importance and Significance of the Research

Archaeological excavations carried out in the ancient site of Iran have traced the traces of the great civilizations that have emerged in this vast area over the millennia. Archaeological evidence suggests that the site has been home to countless generations since ancient times. Assessment of city tourism indicates that important strengths point of tourism in this city having historical monuments, a valuable opportunity for investment in the field of tourism and geographical position can attract more visitors it seems that improving services can attract more people to visit lots of heritage sites in every city in Iran.

## 2. Literature Review

### 2.1 The Silk Road in History

As mentioned earlier, Iran, or the so-called Great Persia, has been one of the main actors in the trading route. On the one hand, silk textiles were fabricated in Iran by getting the materials like silk from the east and then Persians sold the finished products to the westerners. For this reason, some merchants travelled all the way from the Mediterranean to China and returned. The distance between the two regions was ~7000 km (4000 miles), and thus they encountered numerous dangers and hazards on the ways. Therefore, a majority of them travelled short distances to the next market for exchanging their commodities and went back (web, 3). For this reason, the location of Iran has been very important on the trading routes. The term "Silk Roads" was first named in literature in 1877 by Ferdinand Freiherr von Richthofen (1833-1905), a German traveler and geographer, who studied how valued commodities traveled between China and Rome in ancient times. In this regard, Albert Herrmann is the first person who used it in the book title in 1910, which was largely read and cited by the prominent authors of these regions (web, 3).

The Silk Road has significant meaning both for Iran and for the world. Back in its heyday, it was the world's great supply chain and contributed to the extensive development and prosperity of humans for nearly two millennia ; in those days, it performed as one of the main vehicles to improve inter-cultural dialogues and reinforce regional cohesion, peace, as well as solidarity (web, 3), as well as being a great opportunity for culture fusion for countries along the road, and connecting different peoples and cultures in a way that encouraged human creativity, for example, the Renaissance in Europe (Kurin, 2002). The Silk Road was once commended as "if only considering trade volume and number of tourists, the silk road was one of the best in history; the way it could change history, greatly due to people spreading their own culture to other countries via the Silk Road", (Hansen, 2012). During this process, the transportation and communication revolutions not only accelerated the movement of goods but also the transfer of knowledge (Ma, 1996). Nearly, all over the early history, the territory that is known presently as Iran has been called the Persian Empire and the first great dynasty has been the Achaemenid in Iran who governed from 550-330 BC. This dynasty has been established by Cyrus the Great and this era has been followed by the defeat of Alexander the Great from Greece and the Hellenistic period. Owing to Alexander's conquests, the Parthian dynasty ruled for almost 500 years followed by the Sassanian dynasty until 661 AD.

However, skilled sailors travelled between India, China, and the Persian Gulf; therefore, several commodities have been unloaded at the ancient port of Siraf. Consequently, the commodities have been transferred from Siraf to different directions, toward capitals and major trading cities like Susa, Persepolis, Merv, Bukhara, and Babylon. Todays, the remains of the ancient cities could be seen across Iran. Additionally, the northern branch of the secondary route was crossing through the Caspian Sea. The significance of the Royal Road in ancient Persia should not be ignored. According to historians, this route has been the foundation of the silk road, which has been established at the time of Persia's famous king, Darius the Great. Darius has been the 4$^{th}$ king of the Achaemenid dynasty and it is well known that numerous major construction projects have been constructed during his governance. Moreover, the Royal Road has been constructed from Sardis on the west side of current Turkey all the way to Susa in the southwest of Iran for fast accessibility to the furthest end of the empire (Naderi, 2014).

However, the Arabs defeated Iran in the 7$^{th}$ century and presented Islam to the people. Further invasions and occupations were made initially from the Turks and then from the Mongols. In this period, local dynasties like Afsharid, the Zand, the Qajar, and the Pahlavi started in the early 1500s and once again dominated. Finally, the Pahlavi dynasty was defeated by the Islamic revolution in 1979 and thus Iran's government followed Islamic principles (web,4). Here is a brief introduction of ancient dynasties' timeline of Iran:

| CE | |
|---|---|
| 224 | The Sassanid Empire is founded by Ardashir I. |
| 421 | Bahram V becomes king. He will later become the subject of many tales and legends. |
| 661 | The Arabs invade Iran and conquer the Sassanid Empire. They bring the Islamic religion and Islam rule to the region. |
| 819 | The Samanid Empire rules the region. Islam is still the state religion, but the Persian culture is revived. |
| 977 | The Ghaznavid dynasty takes over much of the region. |
| 1037 | The rise of the Seljuq Empire founded by Tughril Beg. |
| 1220 | The Mongols invade Iran after Mongol emissaries are killed. They destroy many cities, killed much of the population, and caused devastation across Iran. |
| 1220 | - The Mongols invade Iran after Mongol emissaries are killed. They destroy many cities, killed much of the population, and caused devastation across Iran. |
| 1350 | The Black Death hits Iran killing around 30% of the population. |
| 1381 | Timur invades and conquers Iran. |
| 1502 | The Safavid Empire is established by Shah Ismail. |
| 1587 | Shah Abbas I the Great becomes king of the Safavid Empire. The empire reaches its peak under his rule becoming a major world power. |
| 1639 | The Safavid Empire agrees to a peace agreement with the Ottoman Empire called the Treaty of Zuhab. |
| 1736 | A weakened Safavid Empire is overthrown by Nadir Shah. |
| 1796 | The Qajar dynasty is established after a civil war |
| 1908 | Oil is discovered. |
| 1921 | Reza Khan captures Tehran and seizes power |
| 1935 | The official name for the country is changed to Iran from Persia. |
| 1941 | A new Shah, Mohammad Reza Pahlavi, is put into power. |
| 1979 | The Islamic Republic of Iran is declared. |

Figure 2: Governors of the ancient Persia according to dynasty (BCE).

Figure 3: The governors of the ancient Persia according to dynasty (CE).

The Silk Road was flourished in the course of the Sasanian era. Actually, a majority of the wealth of Sasanian Empire had been obtained via the exchange of products for silk across the Silk Road. In this way, traders engaged in importing and exporting numerous kinds of products like a variety of carpets, valuable gems and stones, spices, fabrics, as well as each kind of high-consumption products through this ancient route. However, receiving taxes and customs duties from caravans was essential for the Sassanian Empire because they were the largest source of revenue (web, 5). It is widely accepted that King Darius the Great set military check points on the mentioned roads to ensure the caravans' safety. Moreover, the road connecting the city of Shush and Sardis continued from the Silk Road that has been constructed for boosting the silk trade between the West and East. Based on the accounts of the stories proposed on the Silk Road and ancient books, the Romans and the Byzantine Empire have largely attempted for boosting their commerce and enhancing their Empire. For this reason, numerous battles occurred between the Romans and the Iranians for exchanging the products via the Silk Road. On the one hand, the Byzantine Empire had continually attempted for eliminating its dependence on the Iranian people for exchanging considerable volumes of gold and valuable stones via the Silk Road. Several Empires sought for increasing their commerce with China using the Silk Road and thus directly imported silk from China through the Ethiopian commercial ships across the Indian Ocean. On the other hand, the Iranian people were capable of closing the ships' routes in the Red Sea, the Indian Ocean, as well as Yemen. Finally, each Western effort, in particular, the Byzantine Empire efforts, completely failed to dominate the Silk Road (web, 5).

Moreover, ancient times are popular for multiple governments who were in completion with each other to first determine the goods' price. Hence, the governments crucially made policies for dominating the land routes and seas for achieving these objectives, so the Silk Road included these polices. In addition, China has been the focus of the production of silk since ancient times, which has been a major profitable item imported from the East (Naderi, 2014). Actually, we see the development of the economy in the East when the silk produced because the eastern people differently transported silk to various areas. In this way, the military victories of Dariush and Xerxes, the Achaemenian emperors, provided the ground for emerging the Silk Route and thriving the silk business. With the commercial relationships between India and Iran, Central and East Asia involved in exchanging multiple worthwhile commodities using this ancient path. After that, we see the gradual extension of the Silk Road to the Mediterranean Sea. Nonetheless, the greatest economic development in the Silk Road was observed in the course of the Sassanian governance (web,6). Even though trade experienced its expansion in ancient times via the Silk Route and several merchants could be popular, this emerging business could not continue continuously. Over time, this economic ancient road was used less than before so that we do not see anything about those day except for its name. However, numerous parameters contributed to the elimination of the Silk Road. With the emergence of novel technologies, experts in the field tended to the production of specific commodities, including the ones that have been already imported via the Silk Route. Thus, multiple businesses vanished over time. Technological advances in the transport innovation resulted in the construction of a lot of profitable vehicles that considerably declined the contribution of the Silk Road. Particularly, railway industry could link a majority of the areas of the former Soviet Union to each other. Notably, the absence of security, bandits, battles, and other factors in countries' policies have been proposed as some reasons for the elimination of the Silk Road (web, 6).

## 2.2 The Silk Road in the Present

### 2.2.1 UNESCO Programs for Revitalization of the Silk Road

UNESCO began the Integral Study of the Silk Roads: Roads of Dialogue project for highlighting the complicated cultural interactions stemming from encounters across the Silk Roads (Williams,2015), which is an early Silk Roads-related culture study. In 1994, the United Nations World Tourism Organization (UNWTO) promoted Silk Road tourism at the 5th International Meeting in Samarkand, Uzbekistan, 19 countries demanded "...a useful and peaceful re-emergence of such legendary paths as a universal richest cultural tourism destination..." in the Samarkand Declaration, which was the beginning of the Silk Road Action Plan. Moreover, over time, UNWTO cooperated carefully with the main UN agencies like Scientific and Cultural Organization (UNESCO), the United Nations Development Programme (UNDP), as well as the United Nations Educational for advancing sustainable development necessities in regions around the Silk Road (web, 3).Originally this specialized program was related to the tourism expansion along with the Silk Road and in the meeting, three focus areas including promotion and marketing, travel facilitation, destination management, and capacity building. According to the action plan, seven main Silk Road stakeholders were defined as UNWTO secretariat, Silk Road task forces, partner UN agencies, educational institutions, NGOs, Silk Road member states, and other agencies, and UNWTO affiliate members and stakeholders of private sectors (web, 3). Then, 33 Member Countries participate in the UNWTO Silk Road Program, including China. In addition, UNESCO set a 5,000 km stretch of the Silk Road network from Central China to the Zhetysu Region of Central Asia as one of the new World Heritage Sites in 2014.

## 2.2.2 Case Study

As I mentioned before archaeological evidence suggests that Iran has been home to countless generations since ancient times. So I chose one case study along this road in Iran. Kerman- shah is located in the western part of Iran. Kermanshah stems from the Sasanian eratitle Kermanshah that is translated to "King of Kerman". It is clear that the son of Shapur III, Prince Bahram held the title when he was appointed as the governor of the province of Kirman (now is known as Kerman Province). After that, with his succession as Bahram IV after his father in 390 (r. 388–399), he founded Kermanshah, and utilized his earlier title to the new city; that is, City of the King of Kerman" (web, 7). With regard to of its attractive landscapes, antiquity, Neolithic villages, as well as rich culture, Kermanshah was regarded a cradle of prehistoric cultures. Considering the archaeological excavations and surveys, prehistoric people occupied the region of Kermanshah since the Lower Paleolithic era, which was continued to later Paleolithic eras till the late Pleistocene era. In fact, the Lower Paleolithic evidence includes several handaxes observed in the Gakia area to the eastern part of the city and experts found the Middle Paleolithic remains in diverse parts of the province, in particular, in the northern vicinity of the city in Tang-e Malaverd, near Taq-e Bostan, and Tang-e Kenesht (Limbert, 1968).

Furthermore, Neanderthal Man was shown to exist in the Kermanshah area in this era so that just one skeletal remains of this early human discovered in Iran were discovered in rock shelter and three caves located at Kermanshah province. It is notable that Warwasi, Qobeh, Malaverd, and Do-Ashkaft Cave are the popular Paleolithic caves in this region. This area is a first place, wherein the human settlements like Qazanchi, Asiab, Chia Jani, Ganj- Darreh, and Sarab have been founded between 8,000 and 10,000 years ago, when nearly the first potteries of Iran have been fabricated in Ganj-Darreh near the present Harsin (Naderi,2014). With regard to the studies by the University of Hamadan and UCL in May 2009, the head of Archeology Research Center of Iran's Cultural Heritage and Tourism Organization informed that one of the oldest prehistoric village in the Middle East, which dates back to 9800 B.P., has been discovered in Sahneh situated in west of Kermanshah (Sahamieh and Moradpour,2015). It was found that Sassanid rulers selected a pretty context for their rock reliefs in line with a historic Silk Road caravan route waypoint and campground so that these reliefs were close to the holy springs flowing into a big lake at the bottom of a mountain cliff.

### 2.2.3 The Bisotun World Heritage Site

Persian civilization shows itself off on an elegant rocky mountain near Kermanshah, in west- ern Iran, Fig 4. The Bisotun World Heritage Site, dating back to 521 BCE, is situated at Kermanshah Province in the west of Iran (Naderi, 2014).On Mount Bisotun different rulers from the ancient world carved mementos that have become precious legacies to the modern world. Together, these Median, Achaemenid, Parthian, Sassanid, and Safavidgems have created the world's largest rock carving: Behistun Historical Site. Except for Darius the Great's remarkable Bisotun Inscription (UNESCO heritage), Bisotun Historical Site also features Farhad Tarash, Balash Rock Relief, Statue of Hercules, and Shah Abbasi Cara vanserai. Bisotun historical and cultural site is a historical and archaeological site related to different pre-Islamic, post-Islamic, and prehistoric historical periods (web, 8). Bisotun Historical Site includes one of the world's most significant historical documents:

- Bisotun Inscription (UNESCO)
- It includes the world's largest stone inscription
- The site is a window to many historical periods
- It is a charming collection of historical rock reliefs, inscriptions, statues, and structure
- It features the close tie between Persia and Greece

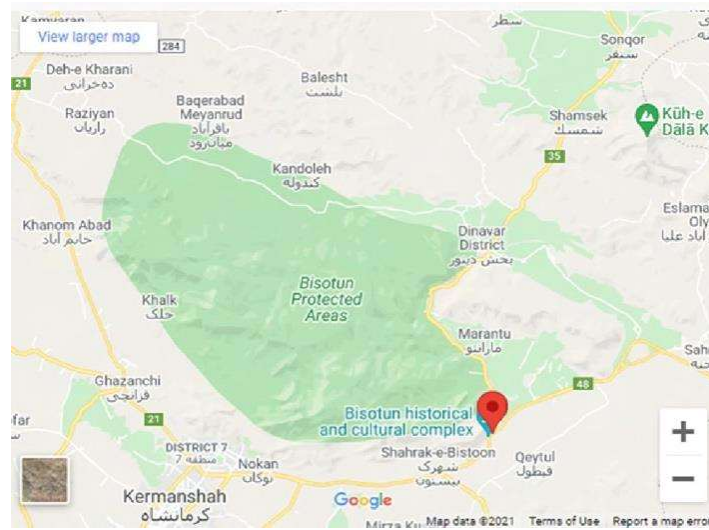

Figure 4: Map of Bisotun Historical Site.

The mountain had become a collage where kings of different historical periods left a valuable relic. This shows the mountain's significance among ancient Iranians. Above all, they probably considered the rocky mountain a sacred place. Plus, the mountain's special location along one of the main routes that connected Persia to Mesopotamia made it a great place for the kings to show off their grandeur, Fig 5. Being next to a vast green field and providing them with a high altitude that preserved their works from destruction might also have been among the reasons why rulers of different eras chose Mount Bisotun and left us with the amazing Bisotun Historical Site. This Mountain has been one of the most important factors in the attractiveness of this area for the construction of historical monuments and is one of the highest walls in Iran and the fifth-highest wall in the world (Mohammadi et al., 2010).

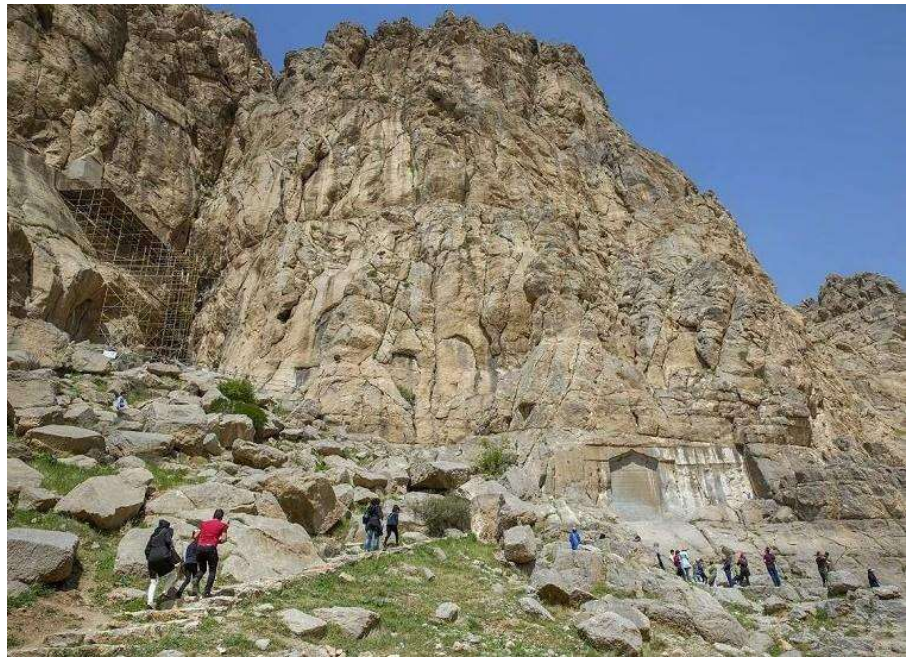

Figure 5: Bisotun Mountain

### 2.2.4 Bisotun inscription

After Cyrus the Great died, a series of events led to the succession of Darius the Great to the throne. When he became the ruler of the vast Achaemenid Empire in 521 BC, he ordered the construction of many monuments. One of them was a bas-relief and cuneiform inscriptions high on the face of

Mount Bisotun (Naderi, 2014). As the most treasured relic in Behistun Historical Site, Behistun Inscription has been considered the only recognized monumental text of the Achaemenids that documents a specific historical event. In the 19th century, the trilingual inscription became a Rosetta Stone: a key to decoding previously undeciphered ancient languages. Bisotun inscriptions are in Old Persian, Elamite, and Babylonian languages. In them, he documented his genealogy, bragged about his royal virtues and power, and recounted the story of how he defeated rebellions and other pretenders to the throne and reestablished the Empire. The inscription is 15 meters high and 25 meters wide (Naderi, 2014). The relief portrays Darius, holding a bow (symbol of power) and stepping on the chest of a pretender to the throne. Two noblemen are standing behind the crowned king. Plus, nine rebel leaders are standing before him with tied hands and ropes around their necks. Above them, you can also find Ahura Mazda (Zoroastrian God) with a circular band in his hand (symbol of the vow between mankind and god) Fig. 6.

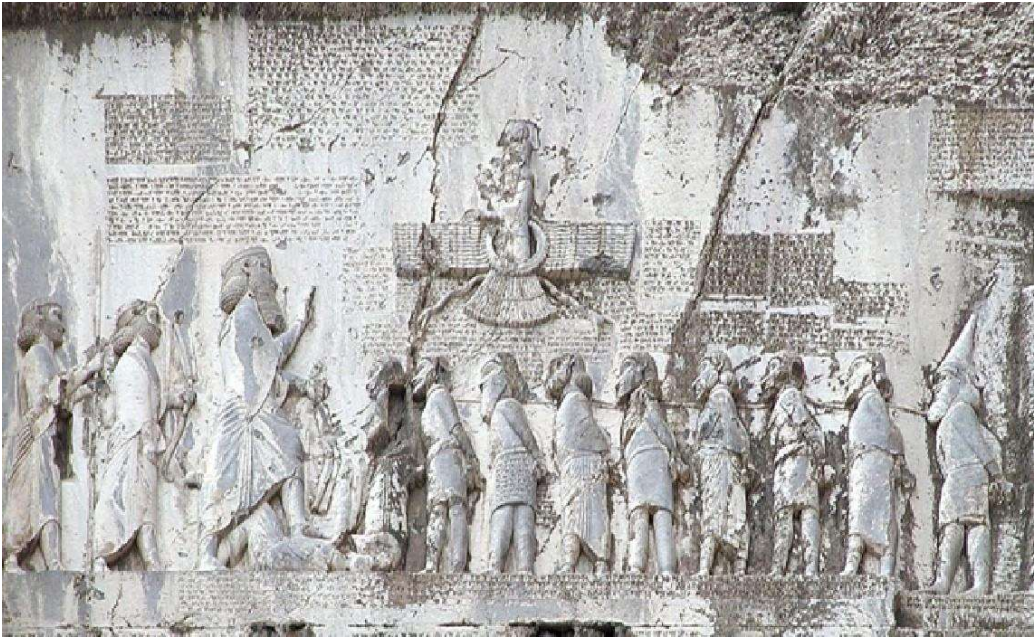

Figure 6: The Bistun inscription.

## 2.2.5  Taq-e Bostan

The historical architectures of all nations reflect their arts, cultures, and histories. Therefore, the

Sassanid inscriptions in diverse locations of Iran exhibit the majesty and distinction of the Sassanid dynasty, Fig. 7. These humans attempted to show off their magnificence and power using relics and inscriptions on the mountains on the leading ways like Silk Road because these places could show all their glory to passersby. For example, Taq-e Bostan in Kermanshah province has been known as a key Sassanid rock relief, which implies the Sassanid kings' power to the future generations. According to the experts, it is one of the exciting places on Traveling to Iran, which deals with the majestic parts of Iran's history that have been carved on the stone (Naderi, 2014).

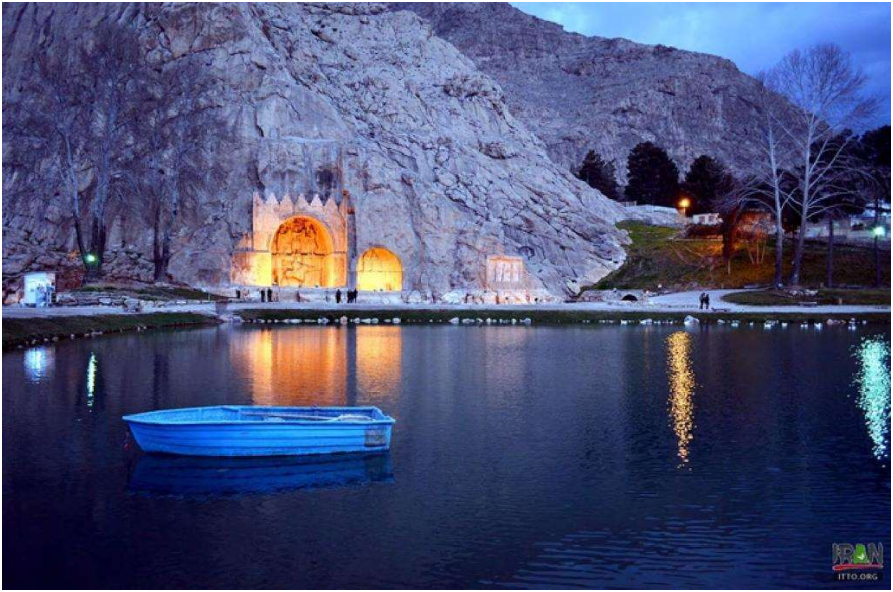

Figure 7: The Sassanid rock relief in Kermanshah known as Taq-e Bostan.

One of the sites with a collection of rock reliefs carved on a mountain in the vicinity of a spring flowing into a pool at the base of the mountain in Kermanshah, which is known as Taq-e Bostan, contains a rock relief and the two rock-cut arches so that one of them (with the height of 9 m) is larger than the other. Based on many Iranologists, this arch dates back to Khosrow II (around the fourth century AD) that provides much information of the religious tendencies, jewels, beliefs, and clothes of the Sassanid era. In fact, the relic demonstrates the coronation of the Sassanid king (Naderi,2014) who is  standing on a platform so that his right hand is stretching toward Ahura Mazda on his right-hand side and his left hand is on his sword. Actually, Ahura Mazda is giving a beribboned ring to the king. Then, Anahita, on his left-hand side, is standing and keeps the water jar in her left hand and another beribboned ring in her right hand, Fig. 8 (Naderi, 2014).

   According to the above discussions, each of the noted relics may provide several evidence of the jewels and clothes employed by the people and kings in the Sassanid era. In this regard, we show an example. The king riding over the horse is wearing colorful clothes that are decorated with woven geometrical shapes and golden threads. And, at the scene of boar hunting, the clothes of the king are decorated with the figure of the Simurgh. Each king portrayed on the reliefs is wearing the earrings and necklace and their clothes contains long loose folded pants, a belt around their waists, as well as a knee-length coat. Moreover, they have bushy eyebrows and hair and a beard. Of course, several things have been carved on their thrones. For example , the entourage's clothes are ornamented with the birds and plants (Mohammadi et al., 2010).

Figure 8: Taq-e Bostan, The king is standing on a platform

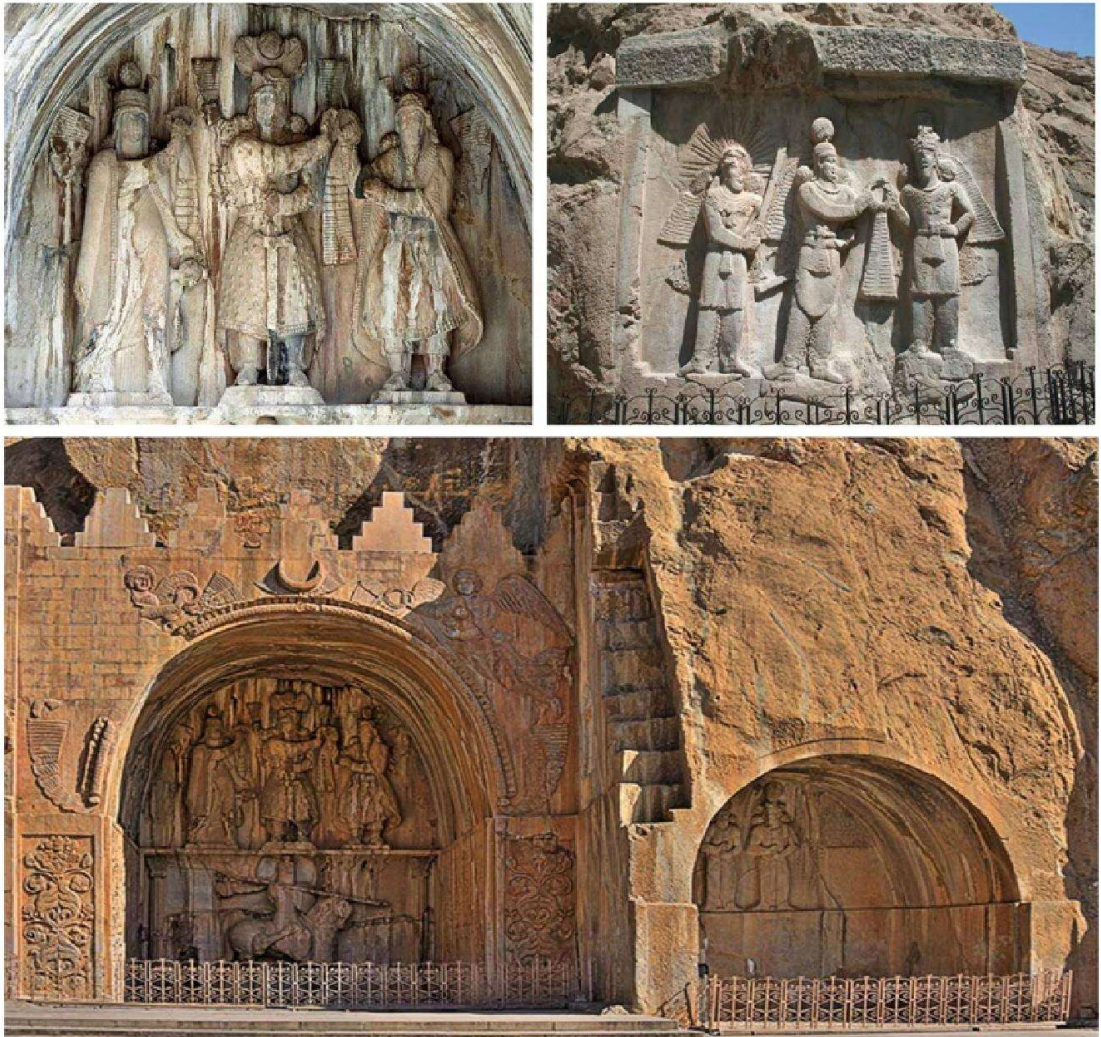

2.2.6 **Bisotun Palace**

This palace is located near the Bisotun inscription and opposite it, somewhere along the ancient road that connects Bisotun Mountain and Biston Lake (Mohammadi et al., 2010). Archaeological excavations show that this building was abandoned partially during the Sas- sanid period, and at that time only the surrounding walls were built. 500 years later, with the construction of sixty-four rooms inside the building, with bricks, it was used as an inn, Fig 9. This caravanserai collapsed in the area after an earthquake and only the Sassanid stones remain. It is included in the UNESCO World Heritage Site of Bisotun (web, 8).

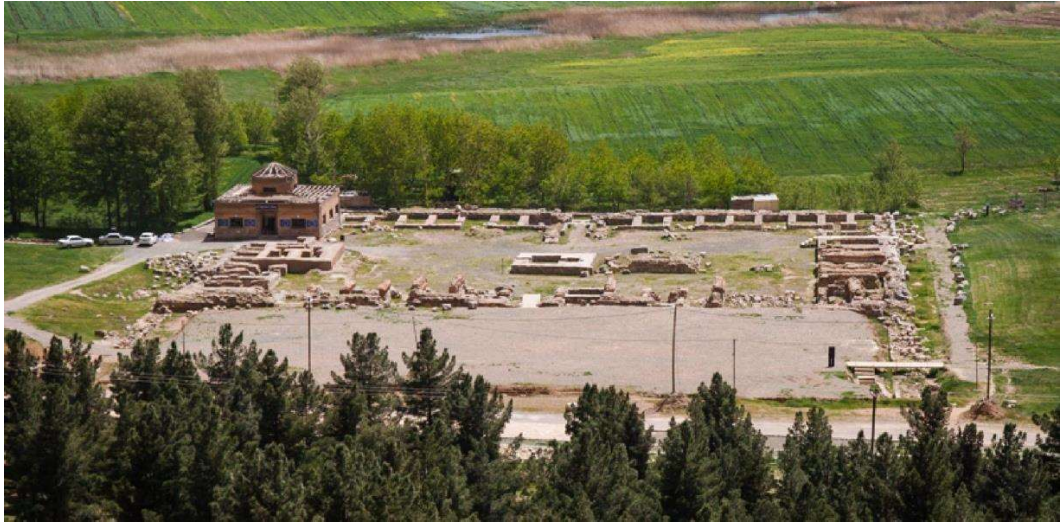
Figure 9: Bisotun Palace.

Countless relics of Iran's precious heritage were either destroyed during the years. For instance about my case study, Kermanshah is home to thousands of tourist attractions, including historical sites. Unfortunately Out of the 4,500 ancient sites and monuments, experts registered 736 cases on the list of National Heritage Sites. So the rest of them need to preserve and registered on UNESCO and National heritage sites (web, 8).

Based on the recent universal tourism tendencies as well as the certain heritage attractions of the province, the government detected the tourism sector as the main economic operator. According to the documents, the first governmental agent in charge of tourism was designated in Iran in 1935 that was popular for the "attraction of tourists and advertisements". After that, this agent experienced numerous changes in its structure, name, policy, and aim (Farzin, 2007). Nevertheless, the imposed Iran-Iraq war and political crises like the Islamic Revolution of Iran in 1978 had unpleasant effects on the Iran tourism industry but tourism industry was recognized as the first development plan of the country after the Islamic revolution (Safaei, 2007).

These days, the Iran Cultural Heritage, Handicraft and Tourism Organization is the government agent in charge of establishing, protecting, preserving, and restoring the country's cultural and historical legacy, and developing and promoting tourism. Hence, the prominent purposes of tourism development in the five years (2004-2009) economic, cultural, and social development plan of Iran are the introduction of the civilization and culture of Islamic Iran, which would result in local industries and commerce, and demonstration of an appropriate global image of Iran(Farzin, 2007). In recent decades, the growth of the 'heritage industry' is becoming obvious especially in the context of tourism .This key feature of heritage tourism makes it a powerful economic- increasing tool. Studies have shown that heritage tourists usually stay longer and spend more money on this travel experience than other kinds of travelers, thus making this form of tourism an important economic development tool. so this heritage site and heritage road have the special capacity for tourists. All these materials can attract visitors to visit and know more about the history of the road and visit most of the cities and countries along this road and all these also can create job opportunities (Safaei, 2007).

# 3 Research Gap

Generally, the study of Silk Road tourism in Iran and especially in Bisotun, deserves the attention of academia. The mainstream literature on Bisotun tourism mainly discusses heritage conservation, the relationships between economic growth and tourism, cultural tourism, tourism destination marketing, etc. Generally speaking, there are very few studies connecting heritage tourism and the Silk Road. Thus there is a gap between the two spheres of heritage tourism and the Silk Road.

# 4 Methodology

This research is being developed by using two interrelated approaches: first, the investigation of primary and secondary resources. Primary resources include photos, digital documentation, paintings, and videos, while secondary sources such as scholarly journals, books, and articles will be consulted. I am proving standards and guidelines for the use of Augmented Reality (AR) in the field and consider a framework to evaluate and use AR for the preservation of cultural heritage. The methodology approach to the research will be designed to examine and validate AR and Virtual Reality (VR) before modifying or demolishing historic structures. Apparently, with regard to Iran's potential for gaining international tourists according to both approaches; that is, cultural–heritage tours and economic cartels, improvement in the services can attract more people to visit lots of heritage sites in every city in Iran.

# 5   Future Research

Recently some museums are using AR as a virtual tourism. Moreover, the expansion of mobile devices like smart phones and tablets, as well as modification for accessibility of the Internet led to changes in the research methods and conservation of heritage and the sharing and teaching ways (Ibáñez-Etxeberria and Kortabitarte,2006; Zandi, 2021). On the other hand, museums could provide the ground for testing the use of such new technologies as educational instruments, but what about heritage sites? My future research is will be using new technology for conserving heritage places. These technologies could qualify buildings, monuments, and spaces that do not exist today and thus cannot be reconstructed. Hence, this may empower us to view an object with its complete or incomplete details without damaging it. In addition, it may show a touristic route that has been promoted with super-imposed information, which illustrates the intended locations. Finally, it may show a virtual tour guide to the visitors, and also these technologies may fascinate and encourage visitors to travel and feel the sense of places.